\documentclass[twocolumn]{aastex631}
\usepackage{amsmath}
\usepackage{mathtools}
\usepackage{color}
\usepackage{hyperref}

\newcommand{\specnotation}[2]{\ensuremath{\rm #1 \, {\scriptstyle #2}}}
\newcommand{\OVI}{\specnotation{O}{VI}}
\newcommand{\atomH}{\specnotation{H}{I}}

\newcommand{\Mstar}{{\ensuremath{M_*}}}
\newcommand{\Mhalo}{{\ensuremath{M_{h}}}}

\newcommand{\CGMsq}{CGM$^2$}

\newcommand{\Msun}{\ensuremath{M_{\odot}}}

\newcommand{\logten}{{\rm log}_{10}}

\newcommand{\NOVI}{\ensuremath{N_{\OVI}}}

\newcommand{\cmmt}{cm\ensuremath{^{-2}}}
\newcommand{\kms}{km s\ensuremath{^{-1}}}
\newcommand{\IP}{\ensuremath{R_{\perp}}}

\newcommand{\rng}{\text{--}}

\graphicspath{{./}{figures/}}
\defcitealias{Tchernyshyov:2022vw}{Paper~I}

\begin{document}

\title{The \CGMsq\ Survey: Quenching and the Transformation of the Circumgalactic Medium}

\author[0000-0003-0789-9939]{Kirill Tchernyshyov}
\affiliation{Department of Astronomy, University of Washington, Seattle, WA 98195, USA}
\author[0000-0002-0355-0134]{Jessica K. Werk}
\affiliation{Department of Astronomy, University of Washington, Seattle, WA 98195, USA}
\author[0000-0003-1980-364X]{Matthew C. Wilde}
\affiliation{Department of Astronomy, University of Washington, Seattle, WA 98195, USA}
\author[0000-0002-7738-6875]{J. Xavier Prochaska}
\affiliation{University of California, Santa Cruz; 1156 High Street, Santa Cruz, CA 95064, USA}
\affiliation{Kavli Institute for the Physics and Mathematics of the Universe (Kavli IPMU) The University of Tokyo; 5-1-5 Kashiwanoha, Kashiwa, 277-8583, Japan}
\affiliation{Division of Science, National Astronomical Observatory of Japan,
2-21-1 Osawa, Mitaka, Tokyo 181-8588, Japan}
\affiliation{Simons Pivot Fellow}
\author[0000-0002-1218-640X]{Todd M. Tripp}
\affiliation{Department of Astronomy, University of Massachusetts, 710 North Pleasant Street, Amherst, MA 01003-9305, USA}
\author[0000-0002-1979-2197]{Joseph N. Burchett}
\affiliation{University of California, Santa Cruz; 1156 High Street, Santa Cruz, CA 95064, USA}
\affiliation{Department of Astronomy, New Mexico State University, PO Box 30001, MSC 4500, Las Cruces, NM 88001, USA}

\author[0000-0002-3120-7173]{Rongmon Bordoloi}
\affiliation{Department of Physics, North Carolina State University, Raleigh, NC 27695-8202, USA}

\author[0000-0002-2591-3792]{J. Christopher Howk}
\affiliation{Department of Physics and Astronomy, University of Notre Dame, Notre Dame, IN 46556}

\author[0000-0001-9158-0829]{Nicolas Lehner}
\affiliation{Department of Physics and Astronomy, University of Notre Dame, Notre Dame, IN 46556}

\author[0000-0002-7893-1054]{John M. O'Meara}
\affil{W. M. Keck Observatory, 65-1120 Mamalahoa Hwy., Kamuela, HI 96743, USA}

\author[0000-0002-1883-4252]{Nicolas Tejos}
\affil{Instituto de F\'isica, Pontificia Universidad Cat\'olica de Valpara\'iso, Casilla 4059, Valpara\'iso, Chile}

\author[0000-0002-7982-412X]{Jason Tumlinson}
\affil{Space Telescope Science Institute, Baltimore, MD, USA}

\correspondingauthor{Kirill Tchernyshyov}
\email{ktcherny@gmail.com}

\begin{abstract}
This study addresses how the incidence rate of strong \OVI\ absorbers in a galaxy's circumgalactic medium (CGM) depends on galaxy mass and, independently, on the amount of star formation in the galaxy.
We use HST/COS absorption spectroscopy of quasars to measure \OVI\ absorption within 400 projected kpc and 300 \kms\ of 52 $\Mstar\sim 10^{10}$ \Msun\ galaxies.
The galaxies have redshifts $0.12<z<0.6$, stellar masses $10^{10.1} < \Mstar < 10^{10.9}$ \Msun, and spectroscopic classifications as star-forming or passive.
We compare the incidence rates of high column density \OVI\ absorption ($\NOVI \geq 10^{14.3}$ \cmmt) near star-forming and passive galaxies in two narrow stellar mass ranges and, separately, in a matched halo mass range.
In all three mass ranges, the \OVI\ covering fraction within 150 kpc is higher around star-forming galaxies than around passive galaxies with greater than $3\sigma$-equivalent statistical significance.
On average, the CGM of  $\Mstar\sim 10^{10}$ \Msun\ star-forming galaxies contains more \OVI\ than the CGM of passive galaxies with the same mass.
This difference is evidence for a CGM transformation that happens together with galaxy quenching and is not driven primarily by halo mass.
\end{abstract}

\section{Introduction} \label{sec:intro}

The circumgalactic medium (CGM) is the extended halo of gas surrounding a galaxy and a key site in the baryon cycle that governs a galaxy's supply of fuel for star formation.
Its physical state mediates the accretion of intergalactic gas and can affect the outcome of feedback processes (i.e., whether winds stall or escape).
The physical state of the CGM is set by the interaction of many factors: radiative cooling, the gravitational potential of the host dark matter halo, energy and momentum injection by feedback, and a variety of other possibly important effects such as cosmic rays and magnetic fields.
Different relative contributions of these factors can yield qualitatively different CGM structures.
A classic example is that when considering gravity and radiative cooling, maintaining a hot, quasi-static CGM inside a stable virial shock requires a sufficiently high halo mass \citep[e.g.][]{Birnboim:2003aa,Dekel:2006aa}.
In the absence of other factors, intergalactic gas can accrete onto the host galaxies of lower mass haloes without shocking \citep[e.g.][]{White:1978aa,Keres:2005aa}.
How the balance of these factors affects and is affected by host galaxy properties is a key question for understanding galaxy evolution.

One part of this question is the role of the CGM in how central galaxies in sub-group scale halos ($\Mhalo\sim 10^{11}\rng10^{12}$ \Msun) quench.
We are still learning how CGM observables differ around star-forming and passive galaxies.
When comparing the CGM between these galaxy classes, it is necessary to control for a number of potentially confounding variables.
CGM properties and observables depend on distance (observationally, on the impact parameter, $\IP$) from a galaxy, galaxy mass (stellar or halo) and color, redshift, environment, and, for some tracers, angular location relative to a galaxy (e.g., \citealt{Bergeron:1986aa,Bahcall:1991aa,Chen:2001aa,Stocke:2006tv,Bordoloi:2011vq,Werk:2013uj,Johnson:2015tj,Tejos:2016aa,Burchett:2016wt}).
At the very least, it is necessary to control for impact parameter and galaxy mass and to restrict comparisons to reasonably narrow redshift ranges.
Comparisons between star-forming and passive galaxies have been done with the necessary controls for cool ($T\sim10^4$ K) metal enriched gas traced by \specnotation{Mg}{II}\ \citep{Bordoloi:2011vq,Lan:2020aa,Anand:2021aa} and for hot ($T\gtrsim10^6$ K) gas traced by X-ray emission \citep{Comparat:2022aa,Chadayammuri:2022aa}.
Star-forming galaxies have higher equivalent widths of cool gas tracers than passive galaxies, but results are still unclear for X-ray emission.

There is a paucity of constraints on gas in the intermediate temperature range, $T\sim 10^{5} K$.
When metal-enriched, gas in this temperature range can be traced by the ion \OVI.
This gas may be near thermal equilibrium \citep{Faerman:2017tg,Voit:2019tv} or it may be part of cooling inflows or outflows \citep{Heckman:2002wz,Bordoloi:2017vr,McQuinn:2018vk,Qu:2018va}.
Some \OVI\ may instead be tracing cool gas ($T\sim 10^{4}$ K) that is diffuse enough for the extragalactic background to photoionize oxygen to \OVI\ \citep[$n_{\rm H} \sim 10^{-5}\rng10^{-4} \; \rm cm^{-3}$, ][]{Tripp:2008tj,Stern:2018to}.
\specnotation{Ne}{VIII}\ measurements strongly suggest the presence of a warm-hot phase \citep{Burchett:2018tn}.

Low redshift star-forming galaxies in general (i.e., without controlling for galaxy mass) have higher detection rates and typical \OVI\ column densities than passive galaxies \citep{Tumlinson:2011wm,Johnson:2015tj,Zahedy:2019vq}.
The only mass-controlled comparison of \OVI\ absorber statistics between star-forming and passive galaxies thus far is found in the supplementary materials of \citet{Tumlinson:2011wm}, but its results are inconclusive.
Because the star-forming galaxies in all other comparisons have lower average masses than the passive galaxies, the interpretation of the difference in \OVI\ absorber statistics is ambiguous: is the \OVI\ dichotomy driven by differences in  halo mass or is it the result of some distinct process associated with quenching?

Which interpretation is correct bears on the connection between feedback, the CGM, and galaxy quenching.
If there is no difference in \OVI\ around star-forming and passive galaxies at fixed mass, then the \OVI\ dichotomy is caused by a change in CGM structure as a function of halo mass and the different halo mass distributions of star-forming and passive galaxies.
This scenario has been suggested in works such as \citet{Oppenheimer:2016wy}, \citet{Fielding:2017ul}, and \citet{Sanchez:2019vc}.
If instead there is a difference at fixed mass, then the \OVI\ dichotomy is caused by some process other than the quasi-hydrostatic evolution of a growing halo.
In one example of such a scenario, integrated active galactic nucleus (AGN) feedback heats and partially drives out the CGM \citep{Mathews:2017uf,Suresh:2017vt,Davies:2020wd,Oppenheimer:2020tx,Terrazas:2020wg,Zinger:2020vm}.
Finding that the \OVI\ dichotomy persists at fixed mass would not automatically mean that quenching disrupts the CGM or some change in the CGM causes quenching: both changes could be caused by a third process, such as interaction with large scale structure.
However, it would mean a more direct connection between the state of a galaxy and its CGM than the scenario where star-forming and passive galaxies of the same mass can have the same CGM.

In this work, we compare \OVI\ incidence rates around star-forming and passive galaxies with stellar masses ($\Mstar$) between $10^{10.1}$ and $10^{10.9}$ \Msun\ controlling for impact parameter and stellar mass and, in a separate comparison, approximately controlling for halo mass (\Mhalo).
For the halo mass comparison, we account for the different clustering properties of star-forming and passive galaxies by using star-formation-dependent stellar-mass-to-halo-mass relations when estimating halo masses.
We build our sample by combining galaxy-\OVI\ absorber pairs from a compilation of new and literature measurements published in \citet{Tchernyshyov:2022vw} (\citetalias{Tchernyshyov:2022vw}) with a small number of additional observations from the CUBS survey \citep{Chen:2020wv}.
The dataset is described in \S \ref{sec:data}.
The mass-controlled comparison between star-forming and passive galaxies is described in \S \ref{sec:analysis}.
We discuss the implications of our findings in \S \ref{sec:discussion} and summarize our results in \S \ref{sec:conclusion}.
We assume a flat-universe $\Lambda$CDM cosmology with $H_0=67.8$ km s$^{-1}$ Mpc$^{-1}$ and $\Omega_m=0.308$ \citep{Planck-Collaboration:2016tp}.
Stellar masses are derived assuming a \citet{Chabrier:2003ta} initial mass function.

\section{Data} \label{sec:data}

We analyze a mass and impact parameter matched set of galaxy-\OVI\ absorber pairs.
Most of the galaxy masses, impact parameters, and associated \OVI\ column densities are taken from the galaxy-absorber data compiled in \citetalias{Tchernyshyov:2022vw}.
This dataset combines measurements from the \CGMsq\ survey \citep{Wilde:2021vr} and the literature \citep{Werk:2013uj,Johnson:2015tj,Keeney:2018wp,Zahedy:2019vq}.
We also include three galaxy-absorber pairs from the CUBS survey \citep{Chen:2020wv,Cooper:2021vu,Boettcher:2021va,Cooper:2021vu}.
In cases where information on galaxy environment is available, we exclude galaxies from our analysis if they are within 1 Mpc and 600 \kms\ of a more massive galaxy.

We do not use the galaxy classifications from \citetalias{Tchernyshyov:2022vw}, which were based on fitting galaxy templates to photometric measurements.
Instead, we use spectroscopic classifications.
For galaxies from \citet{Johnson:2015tj} and the CUBS survey, we adopt the classifications given in those works.
For the remainder of the sample, we make our own spectroscopic classifications.

We use three classification criteria: H$\alpha$ emission equivalent width, H$\beta$ emission equivalent width, and the 4000\AA\ decrement $D_{4000}$\footnote{We use the 100\AA-wide interval definition of $D_{4000}$, with intervals 3850\rng3950\AA\ and 4000\rng4100\AA\ \citep{Balogh:1999vi}.}.
We use multiple criteria because the galaxy spectra were taken with different instruments and cover different galaxy restframe wavelength ranges.
We measure H$\alpha$ and H$\beta$ equivalent widths by fitting the galaxy spectra with superpositions of stellar templates and emission lines.
Fitting is done with \texttt{pPXF} \citep{Cappellari:2017ty} using the MILES stellar library \citep{Falcon-Barroso:2011tz}.
We measure $D_{4000}$ by integrating the galaxy spectra over the appropriate interval and taking the ratio of the results.

Not all quantities are measurable from all spectra.
Of the ones that are measured from a spectrum, we adopt the classification according to the \emph{most preferred} quantity, where the order of preference is H$\alpha$, then H$\beta$, then $D_{4000}$.
A galaxy is classified as star-forming if it has an H$\alpha$ equivalent width greater than 6\AA, a H$\beta$ equivalent width greater than 2\AA, or $D_{4000}$ less than 1.6 \citep{Kauffmann:2003vn,Sanchez:2014ux}.
The quantities usually agree on a galaxy's classification.
Taken pairwise, H$\alpha$ and H$\beta$, H$\alpha$ and $D_{4000}$, and H$\beta$ and $D_{4000}$ agree in 42/42, 34/36, and 41/49 instances, respectively.

\begin{figure*}
    \centering
    \includegraphics[width=\linewidth]{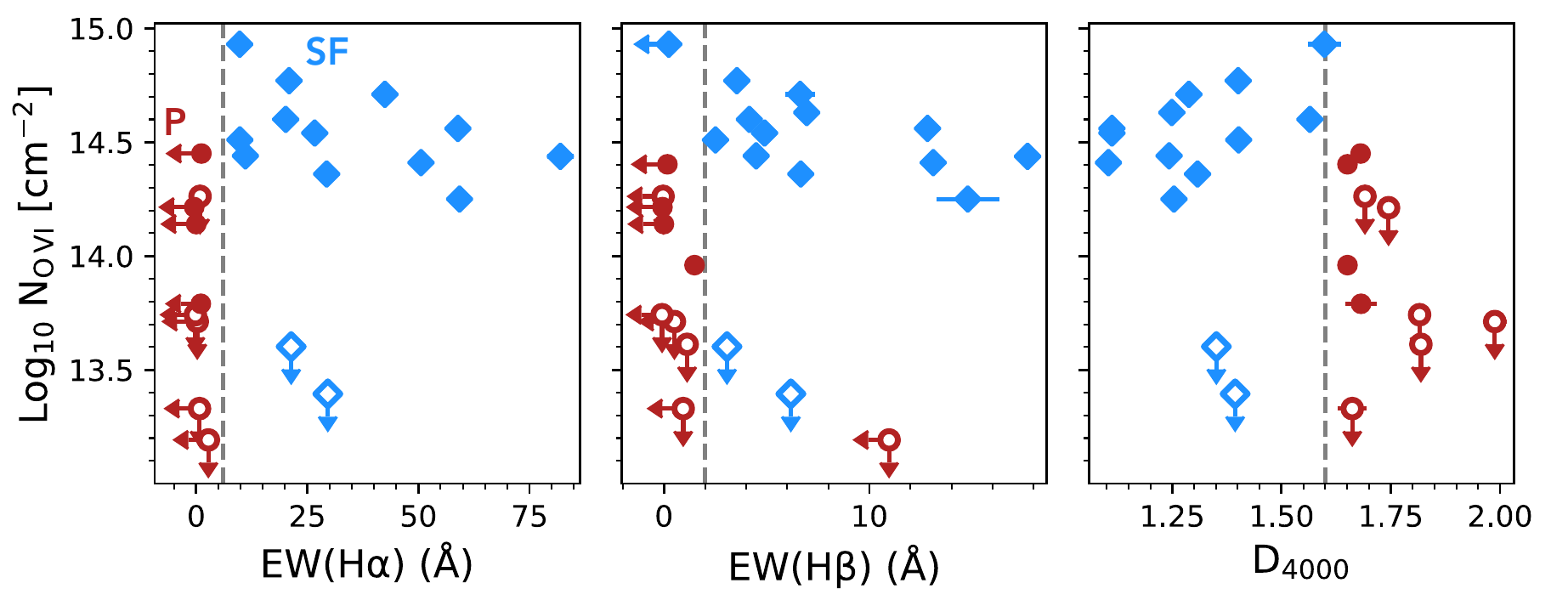}
    \caption{\OVI\ column density as a function of three star formation indicators for galaxies with $\IP < 200$ kpc. Data point colors and shapes denote whether a galaxy is classified as star-forming (blue diamonds) or passive (red and red-outlined circles). Data points that are outlined rather than filled are upper limits on \OVI\ non-detections. Because not all galaxies have measurements of all three indicators, some points only appear in some of the panels. Vertical dashed gray lines denote thresholds used for galaxy classification and were taken from the literature. The three star formation indicators almost always agree on a galaxy's class. There is a clear change in the $\NOVI$\ distribution from one side of a classification threshold to the other.}
    \label{fig:indicators-vs-N}
\end{figure*}

The star formation indicator cuts split the galaxies at low impact parameters into mostly low and mostly high \NOVI\ subsamples.
\autoref{fig:indicators-vs-N} shows \NOVI\ as a function of the three indicators around galaxies with $\IP<200$ kpc.
The two star-forming $\NOVI$\ non-detections have larger impact parameters than all but one of the other star-forming galaxies shown.
Apart form these two non-detections, galaxies on the star-forming side of the classification thresholds have higher $\NOVI$\ than most passive galaxies.

\begin{figure}
    \centering
    \includegraphics[width=\linewidth]{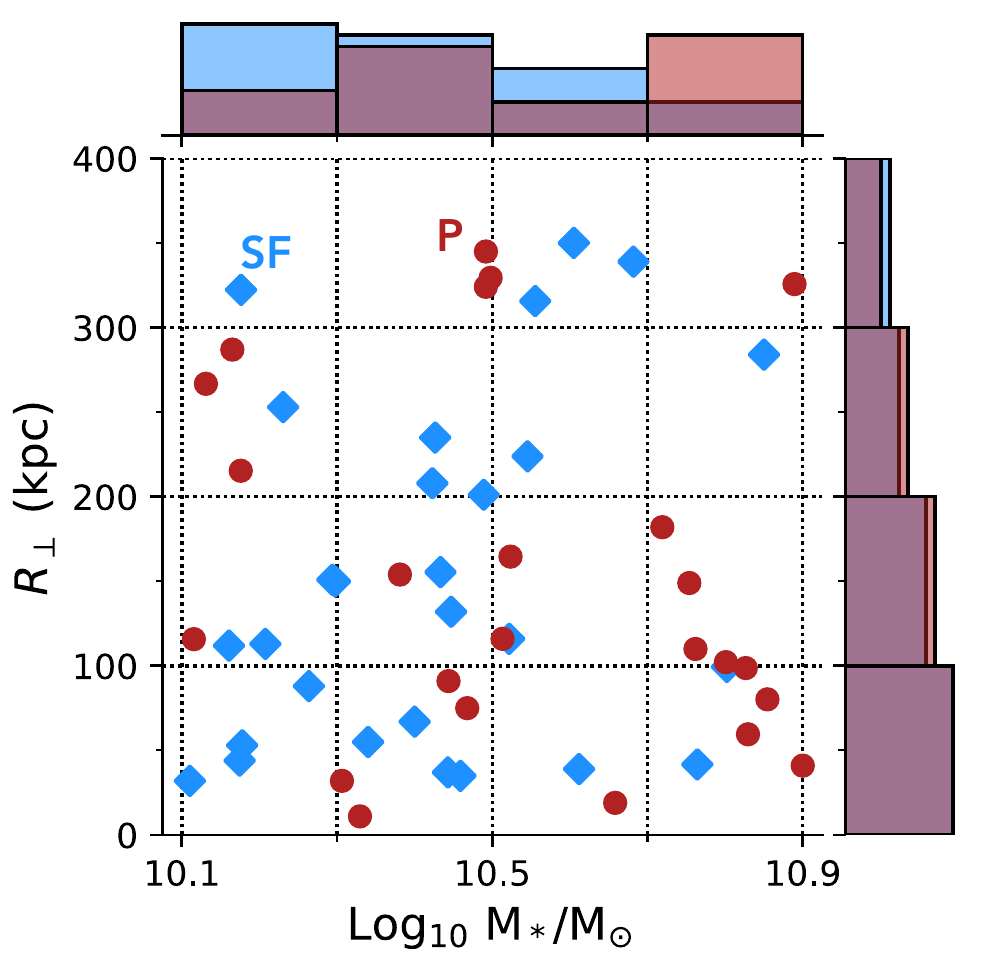}
    \caption{Stellar masses and impact parameters of galaxies in the sample. Blue diamonds represent star-forming galaxies, red circles represent passive galaxies. The distribution of the two classes in stellar mass is shown by the histogram above the scatter plot.}
    \label{fig:rho-mstar}
\end{figure}

We focus on galaxies with stellar masses between $10^{10.1}$ and $10^{10.9}$ \Msun.
\citetalias{Tchernyshyov:2022vw} included galaxies with masses ranging from $10^{7.8}$ to $10^{11.2}$ \Msun.
In that sample, the overwhelming majority of photometrically-classified passive galaxies have $\Mstar\geq 10^{10}$ \Msun.
For this work, we initially spectroscopically classified all of those galaxies with $\Mstar\geq 10^{10}$ \Msun.
The sample contains no passive galaxies with mass less than $10^{10.1}$ \Msun\ and no star-forming galaxies with mass greater than $10^{10.9}$ \Msun.
To get a closer match in $\Mstar$, we restrict the sample to bracket this mass range.
The stellar masses and impact parameters of galaxies in this range are shown in \autoref{fig:rho-mstar}.

To enable a test of the hypothesis that differences between star-forming and passive galaxies are driven by halo mass, we estimate  halo masses for the galaxies in the sample.
We match the observed galaxies to central galaxies from the \texttt{UNIVERSEMACHINE} mock catalogs \citep{Behroozi:2019up} conditioning on star-formation class, stellar mass, and redshift.
We classify mock galaxies to be star-forming or passive using a $10^{-11}$ yr$^{-1}$ sSFR cut, use a tolerance of $\pm 0.01$ dex for matching on stellar mass, and take the nearest redshift available (in all cases, $\vert \delta z \vert \leq 0.1$).
This procedure yields between 800 and 2800 matching mock galaxies per observed galaxy, each with a halo mass.
We summarize an observed galaxy's possible halo mass distribution by its 16th, 50th, and 84th percentiles.
The median halo masses of the star-forming and passive samples overlap for $\Mhalo \approx 10^{11.6}\rng10^{12.2}$ $M_\odot$.
With this halo mass estimation procedure, star-forming and passive galaxies in this halo mass range have $\Mstar = 10^{10.2}\rng10^{10.9}$ $M_\odot$ and $\Mstar = 10^{10.1}\rng10^{10.7}$ $M_\odot$, respectively.

\begin{figure*}
    \centering
    \includegraphics[width=\linewidth]{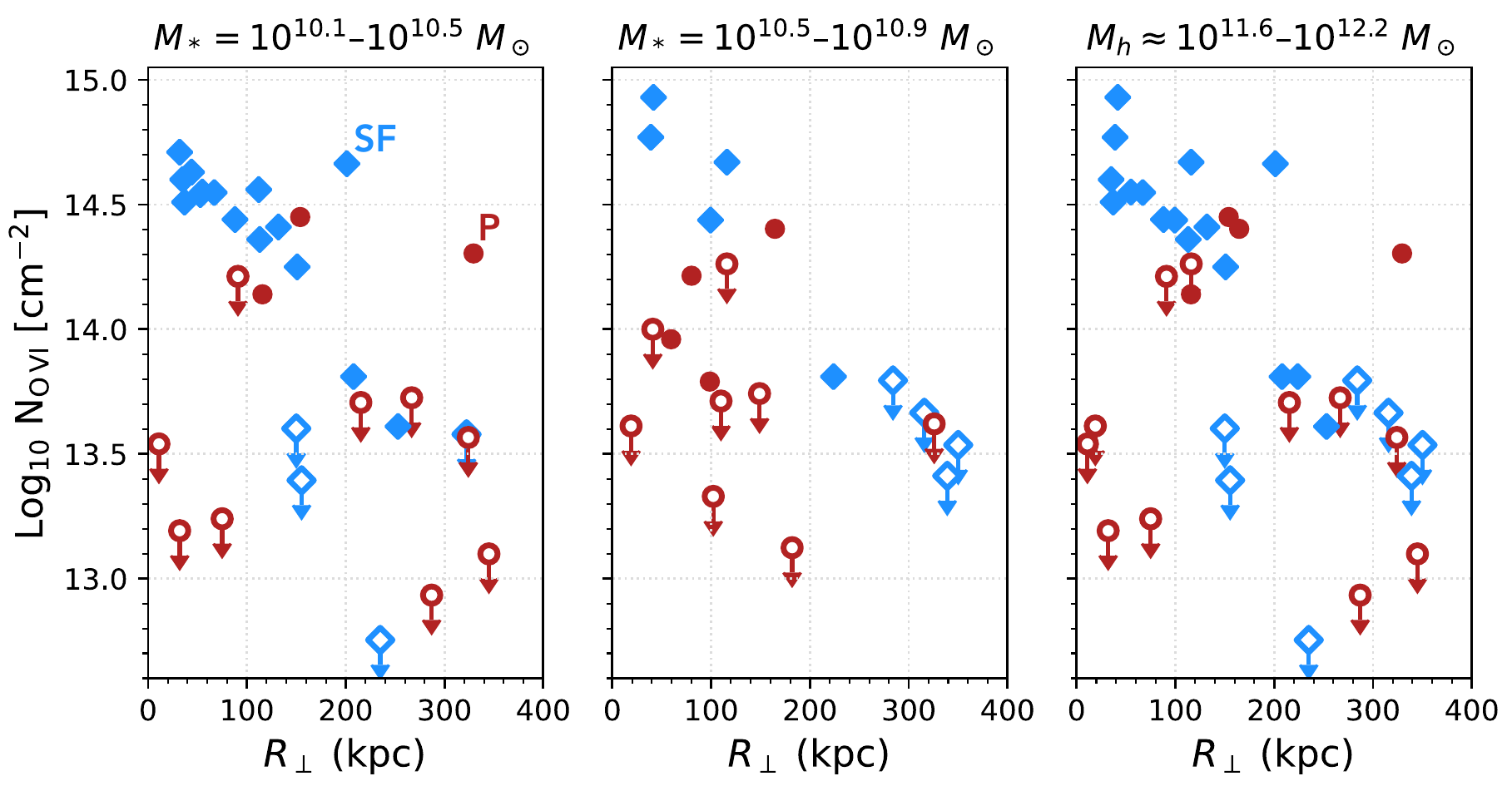}
    \caption{\OVI\ column density as a function of impact parameter for three mass-selected star-forming and passive galaxy comparison samples. Filled data points mark \OVI\ detections while outlined data points mark upper limits on non-detections. Blue diamonds are star forming galaxies and red circles are passive galaxies. The left and middle panels show galaxies selected to be in the same stellar mass range. The right panel shows galaxies selected to be in the same halo mass range, where halo masses are estimated using separate stellar mass-halo mass relations for the two galaxy classes. The halo mass selected sample overlaps with both stellar mass selected samples.}
    \label{fig:rho-N}
\end{figure*}

We compare star-forming and passive galaxies in three mass sub-samples: $\Mstar=10^{10.1}\rng10^{10.5}$ \Msun\ (lower stellar mass), $\Mstar=10^{10.5}\rng10^{10.9}$ \Msun\ (higher stellar mass), and $\Mhalo\approx 10^{11.6} \rng 10^{12.2}$ \Msun (matched halo mass).
The matched halo mass sub-sample partially overlaps with each of the matched stellar mass sub-samples.
The impact parameters and \OVI\ column densities of galaxies in these sub-samples are shown in \autoref{fig:rho-N}.
The right panel shows a subset of the galaxies in the left and middle panels.
The distributions of measurements in all three panels are qualitatively similar.
Star-forming galaxies have distinct inner and outer column density regimes, with uniformly high column densities in the inner region and a broad, generally lower distribution of column densities in the outer region.
Passive galaxies have a broad distribution of column densities at all impact parameters, with a possible tendency towards higher column densities at low impact parameters.
The inner star-forming galaxy column densities are greater than almost all of the inner passive galaxy column densities.

\section{Analysis and Results} \label{sec:analysis}

We quantify the incidence of strong \OVI\ absorbers around star-forming and passive galaxies by calculating covering fractions, the number of detections above a threshold (``hits") over the number of observations.
We adopt a detection threshold of $\NOVI=10^{14.3}$ \cmmt, which is just above the least constraining upper limit in the sample.
An upper limit that is greater than the threshold is ambiguous, meaning that using a lower threshold would require discarding part of the sample.

The number of hits in a sample of fixed size given a covering fraction has a binomial distribution.
Assuming a beta distribution prior on the covering fraction, the posterior probability distribution for the covering fraction is itself a beta distribution.
We use the Jeffreys prior, Beta$(f_C; \alpha=1/2, \beta=1/2)$, and use the 16th and 84th quantiles of the covering fraction posterior probability distribution as a 68\% (1$\sigma$-equivalent) credible interval.

\begin{figure*}
    \centering
    \includegraphics[width=\linewidth]{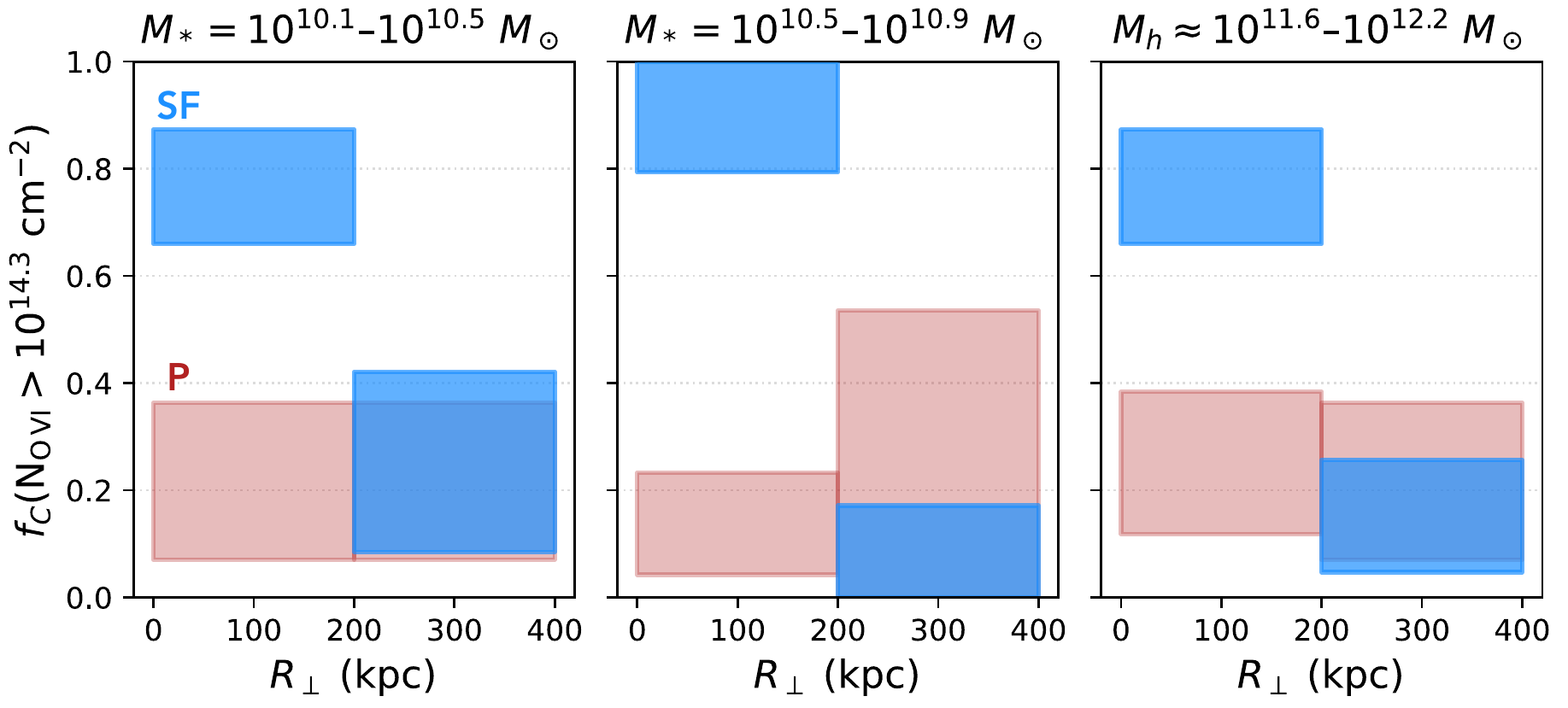}
    \caption{\OVI\ covering fractions of star-forming (shown in blue) and passive (shown in red) galaxies in three mass-selected samples. Shaded regions are 68\% credible intervals about the median for the covering fraction of absorbers with $\IP=0\rng200$ kpc and $200\rng 400$ kpc. Star-forming galaxies have higher inner covering fractions than passive galaxies, but similar outer covering fractions.}
    \label{fig:covering-fracs}
\end{figure*}

From visual inspection of \autoref{fig:rho-N}, there is an obvious need for covering fractions to depend on impact parameter.
We include an impact parameter dependence by splitting each mass-selected sub-sample into inner and outer regions.
\autoref{fig:covering-fracs} shows inner and outer covering fractions for the three mass sub-samples and a dividing $\IP$\ of 200 kpc.
The inner covering fraction around star-forming galaxies is higher than that around passive galaxies.
The outer covering fractions of the two galaxy classes are consistent, with overlapping 68\% credible intervals.

\begin{figure*}
    \centering
    \includegraphics[width=\linewidth]{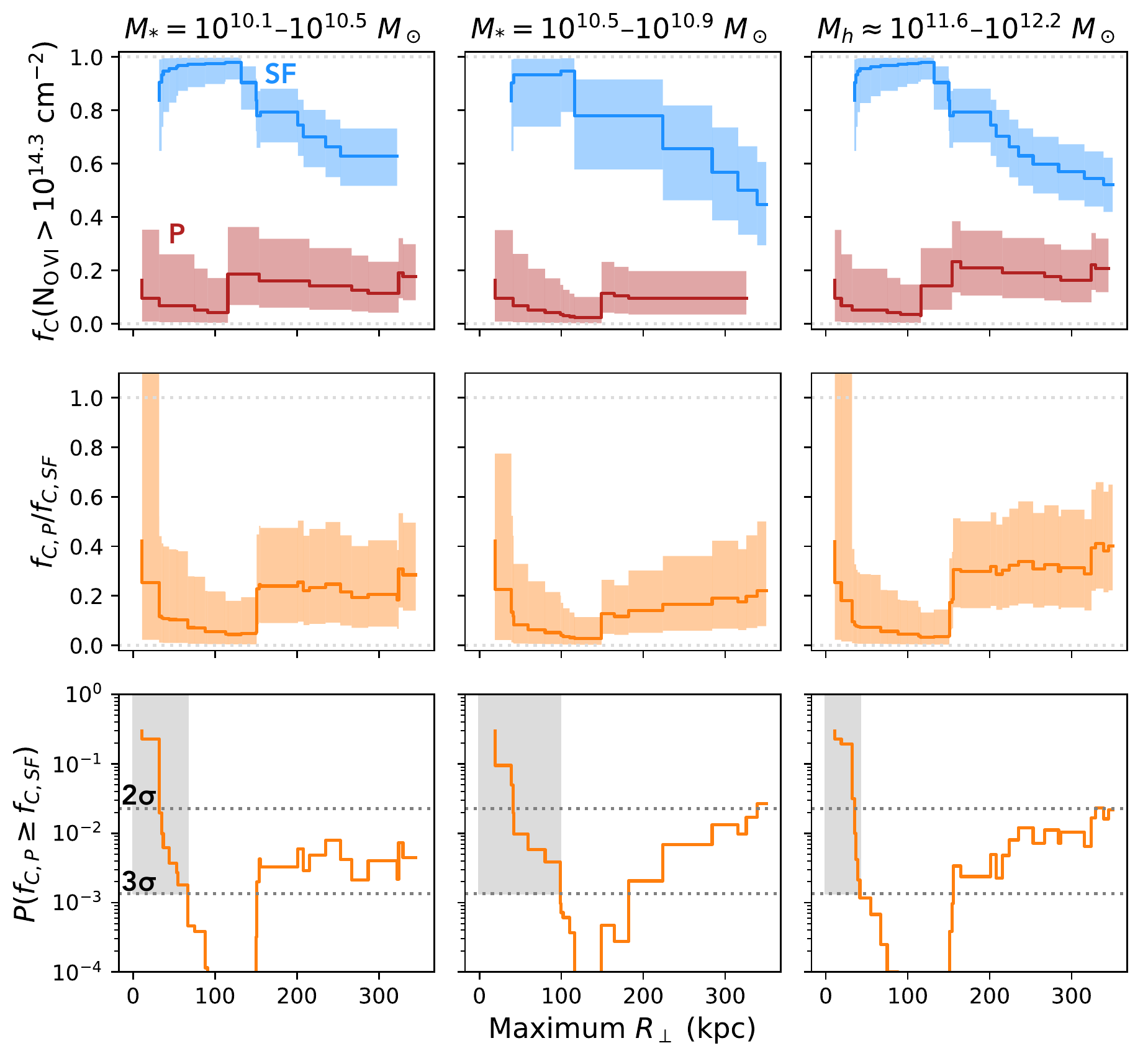}
    \caption{(top row) \OVI\ covering fractions within a maximum impact parameter of star-forming (shown in blue) and passive (shown in red) galaxies in three mass-selected samples. In this row and the next, solid lines are medians and shaded regions are central 68\% credible intervals. (middle row) Ratio of the passive galaxy covering fraction to the star-forming galaxy covering fraction. (bottom row) Probability that the passive galaxy covering fraction is greater than or equal to the star-forming galaxy covering fraction. One-sided $2\sigma$- and $3\sigma$-equivalent probabilities are indicated by horizontal lines. Maximum impact parameters where the probability can never be less than $3\sigma$-equivalent because there are too few measurements are shaded in gray. The star-forming galaxy covering fraction declines for maximum impact parameters greater than approximately 150 kpc. Between the region with too few measurements and 150 kpc, the star-forming galaxy covering fraction is greater than the passive galaxy covering fraction with conclusive, greater than $3\sigma$-equivalent statistical significance.}
    \label{fig:inner-cfs}
\end{figure*}

To quantitatively compare the inner covering fractions for the two galaxy classes, we calculate their ratio as a function of maximum inner impact parameter.
The top row of \autoref{fig:inner-cfs} shows inner covering fractions at all impact parameters spanned by each mass-selected sub-sample.
These covering fractions are cumulative: measurements used to calculate the covering fraction at $R_{\perp,1}$ are also used to calculate the covering fraction at $R_{\perp,2}>R_{\perp,1}$.
The star-forming galaxy covering fractions decline outside about 150 kpc.
The passive galaxy covering fractions are similar at all impact parameters.
The middle row shows the ratio $f_{C,P}/f_{C,SF}$.
The probability distribution over the ratio is estimated by integrating over the joint distribution of the two covering fractions, $p(f_{C,P}, f_{C,SF})$, over lines of fixed ratio.

The bottom row of the figure shows the probability that this ratio is greater than one.
We consider a probability less than about 0.00135 to be statistically significant evidence\footnote{This is the probability of drawing a value that is lower than $\mu-3\sigma$ from a Gaussian distribution.} that $f_{C,SF}$ is greater than $f_{C,P}$.
If there are too few measurements, the probability can never be below this threshold.
For all maximum impact parameters that are less than about 150 kpc and where there are enough measurements, the probability is below the threshold.
The decrease in significance for greater dividing impact parameters is likely physical and can be explained by the decline in the star-forming galaxy covering fraction.
In all three mass-matched sub-samples, there are significantly more strong \OVI\ absorbers near star-forming galaxies than near passive galaxies.
{\emph{These results show that there is a statistically significant dichotomy in \OVI\ around star-forming and passive galaxies at fixed stellar mass and at fixed halo mass.}}

\section{Discussion} \label{sec:discussion}

\subsection{Comparison with previous work on \OVI\ and star formation}

The present study builds on \citet{Tumlinson:2011wm}, which found that there is a higher incidence rate of strong \OVI\ absorption around star-forming galaxies than around passive galaxies at $0.1 < z < 0.4$.
The supplementary material of \citet{Tumlinson:2011wm} contains a comparison of \OVI\ detection rates between the galaxy classes restricted to a mass range where star-forming and passive galaxies in their overlap, $\Mstar > 10^{10.5}$ \Msun.
They find that with $2.6\sigma$-level significance, the detection rate is higher for star-forming galaxies, a suggestive result that motivated our current work.

Like a number of other observational studies \citep{Johnson:2015tj,Zahedy:2019vq}, we confirm the general finding that there is an \OVI\ dichotomy between star-forming and passive galaxies.
We extend the result by establishing with high statistical significance that the dichotomy persists when controlling for stellar mass or halo mass.
The difference in \OVI\ incidence around passive galaxies is evidence for a CGM transformation that is associated with how central galaxies in this mass range quench.

\subsection{Galaxy quenching and CGM transformation}
\label{sec:discussion:physical-picture}

Recent theory and analyses of cosmological hydrodynamic simulations offer a candidate for the required CGM transformation: the heating and ejection of CGM gas by integrated black hole feedback \citep{Mathews:2017uf,Suresh:2017vt,Davies:2020wd,Terrazas:2020wg,Zinger:2020vm,Oppenheimer:2020tx}.
In the \emph{EAGLE} and Illustris-TNG simulations, galaxies quench when the integrated amount of black hole feedback of the appropriate type exceeds the binding energy of the CGM and ejects some fraction of it from the halo.
The bulk of the gas remaining in the CGM after quenching is hotter and more diffuse than before, and as a result has a long cooling time.
\citet{Nelson:2018wy} find that in Illustris-TNG, the \OVI\ mass also drops once a galaxy quenches.
This galaxy quenching mechanism is consistent with our observations (and some observational studies of central galaxy quenching, e.g., \citealt{Reines:2015uj,Piotrowska:2022aa}) because its onset is determined by black hole mass, rather than by galaxy stellar mass or halo mass.
Other quenching mechanisms, such as those in which the interaction of a halo with large scale structure cuts off the supply of intergalactic gas to the central galaxy \citep{Aragon-Calvo:2019ty,Winkel:2021aa}, would also be consistent.

\subsubsection{High passive galaxy \NOVI: geometric or temporal variation?}
While most absorbers associated with passive galaxies have low \OVI\ column densities, this is not universal.
Of the sixteen passive galaxies in our sample with $\IP<200$ kpc, two have $\NOVI$\ values typical of star-forming galaxies.
Put another way, the distribution $p(\NOVI)$ for sightlines near passive galaxies is shifted to lower $\NOVI$\ compared to $p(\NOVI)$ for star-forming galaxies, but does have a high $\NOVI$\ tail.

The simplest explanations for these two cases is misclassification or interloping absorption.
Both galaxies have secure classifications, with very low hydrogen emission equivalent widths and $D_{4000}$ values that are greater than the threshold of 1.6.
Their $D_{4000}$ values are less than the median for the passive galaxy sample, so there is a possibility that these galaxies quenched relatively recently.
Interloping absorption can be significant: \citet{Ho:2021we} find that in the \emph{EAGLE} simulations, the column density of \OVI\ absorbers within $\pm 300$ \kms\ of a galaxy can be twice that of absorbers with radius less than $R_{vir}$.
This could be the explanation for one of the galaxies, where we do not have good info on environment.
The other galaxy is found in a \CGMsq\ quasar field with spectroscopy that is complete to a $g$-band magnitude of 22 within 600 kpc, a depth sufficient to detect galaxies down to $\Mstar \sim 10^{9.5}$ \Msun.
No other galaxies are detected within 600 kpc of the sightline and 600 \kms\ of the passive galaxy in question, suggesting that the high \OVI\ column density is not attributable to an interloping galaxy's CGM.

Possible physical explanations for the the high $\NOVI$\ tail include a patchy \OVI\ distribution and a lag between quenching and CGM transformation.
If \OVI\ around a typical passive galaxy is found in localized, anisotropically distributed structures, the low sample-averaged \OVI\ incidence rate would reflect a low per-galaxy \OVI\ covering fraction.
However, these structures would need to have total \OVI\ column densities close to those of sightlines through a star-forming galaxy's CGM.
A different explanation is that the CGM transformation as seen in \OVI\ starts later or takes longer than quenching.
In this interpretation, the passive galaxies with high $\NOVI$ column densities are ones that quenched recently.

Examples of both of these options, within-halo spatial and between-halo temporal variation, have been seen in cosmological hydrodynamic simulations.
\citet{Nelson:2018wy} find that in the TNG100 simulation, the \OVI\ distribution is smooth and approximately isotropic around star-forming galaxies but patchy and obviously anisotropic around passive galaxies.
\citet{Oppenheimer:2020tx} analyze the evolution of black hole mass, star formation rate, and the CGM in the \emph{EAGLE} simulation.
They find that a galaxy quenches at approximately the same time as the most rapid phase of growth for the galaxy's black hole.
The fraction of total halo mass found as gas in a galaxy's CGM starts to sharply decline at the same time, but the covering fractions of ion absorption, including \OVI, do not drop until about 1 Gyr after quenching.
\citet{Appleby:2022wl} compare CGM properties between star-forming, passive, and intermediate (``green valley") galaxies from the \texttt{SIMBA} simulation.
They define $z=0$ galaxies to be in the green valley if their sSFR is between $10^{-10.8}$ and $10^{-11.8}$ yr$^{-1}$.
Based on studies of the correlation between sSFR and $D_{4000}$, galaxies with these sSFRs are more likely than not to have $D_{4000}>1.6$ \citep[e.g.][]{Bluck:2020we,Angthopo:2020aa}, meaning that we would  classify most of them as being passive.
\citet{Appleby:2022wl} find that these galaxies have \OVI\ covering fractions that are intermediate between those of galaxies with lower or higher sSFR, which would be another example of between-halo temporal variation.

While all three of these simulations show a decline in the incidence rate of \OVI\ absorption as a result of integrated black hole feedback, they implement the feedback in different ways \citep{Schaye:2015aa,Weinberger:2017aa,Dave:2019aa}.
These implementation differences lead to differences in bulk properties such as CGM mass as a fraction of halo mass \citep[e.g.][]{Davies:2020ut,Tillman:2022aa} and could mean that different mechanisms are responsible for the reduction in \OVI\ incidence rates in each simulation.
It is not obvious whether simulation resolution has a noticeable effect on the comparisons we make above because \OVI\ covering fractions are a relatively coarse CGM property.
Some works find that refinement beyond the resolution of these three simulations ($m_{gas} \sim10^6\rng 10^7$ \Msun) does not substantially change coarse CGM properties \citep{van-de-Voort:2019aa,Peeples:2019aa,Suresh:2019aa}.
Conversely, \citet{Hummels:2019aa} find that using finer resolution can reduce \OVI\ column densities throughout the CGM and \citet{Lochhaas:2021aa} find that resolution affects what fraction of energy in CGM gas is thermal rather that non-thermal.

\subsubsection{The case of M31}

Our observations consist of a single sightline per galaxy, and so cannot distinguish within-halo variation from between-halo variation.
The CGM of the Milky Way's nearest massive neighbor, M31, has been measured along multiple sightlines \citep{Lehner:2015aa,Lehner:2020tk}.
We would classify M31 as passive and slightly more massive than the most massive galaxies we consider\footnote{The nominal $\logten \Mstar/\Msun$ is 10.93, where we have applied a Kroupa to Chabrier initial mass function conversion factor of $1/1.06$ \citep{Zahid:2012wz}. The sSFR is $7 \times 10^{-12}$ yr$^{-1}$ \citep{Lewis:2015wl,Williams:2017wg}.}.
\citet{Lehner:2020tk} measure \OVI\ along eight sightlines with $\IP<400$ kpc, with four of these eight sightlines at $\IP<250$ kpc.
$4/4$ and $7/8$ \NOVI\ measurements within 250 and 400 kpc, respectively, are greater than $10^{14.3}$ \cmmt.
In our higher stellar mass passive galaxy sub-sample, the corresponding hit rates are $1/11$ and $1/12$.
Using Barnard's exact test for 2-by-2 contingency tables \citep{Barnard:1947aa}, the probability of M31 having a covering fraction that is less than or equal to that of the $\Mstar = 10^{10.5}$ to $10^{10.9}$ \Msun\ passive sample is less than 0.1\%, a greater-than-3$\sigma$ tension.
We propose three (non-mutually exclusive) ways of resolving this tension: a coincidence of observations with a coherent spatial feature; \OVI\ variation being a between-halo temporal effect; and the possibility of different modes of quenching.

\citet{Lehner:2020tk} find that ions other than \OVI\ with $\IP>R_{vir}$ tend to be detected in a particular direction relative to M31 and suggest that this absorption may be arising in an accreting IGM filament.
The \OVI\ sightlines are in this direction, and so could also be related to this hypothetical filament.
Large structures between the Milky Way and M31 are found in simulations of Local Group analogues \citep{Nuza:2014aa,Damle:2022aa} and a bridge of hot $\sim 2\times 10^6$ K gas between the galaxies has been detected in X-ray emission \citep{Qu:2021aa}.

The second possibility is that M31 quenched recently and the CGM transformation is not yet apparent in UV-accessible CGM tracers.
\citet{Williams:2017wg} find that M31 had a burst of star formation about 2 Gyr ago and \citet{Lewis:2015wl} find that the star formation rate has been mostly declining for the past 400 Myr.
M31 could therefore still be in the period between quenching and a drop in metal absorber incidence.

Finally, it is possible that not all quenching is accompanied by a CGM transformation.
Unlike many passive galaxies, M31 has spiral structure and a substantial gaseous disk ($M_{\atomH} = 5\times 10^{9}$ \Msun; \citealt{Carignan:2006tj}).
The $K_S$ band luminosity of M31 is $L_K = 5 \times 10^{10}$ $L_{\odot}$ \citep{Huchra:2012aa,Willmer:2018aa} and its $M_{\atomH}/L_K$ is about $1/10$.
This \atomH-mass-to-luminosity ratio is greater than that of $\approx 97$\% of early type galaxies in the ATLAS$^{3D}$ survey, but is typical for a low redshift massive spiral galaxy \citep{Serra:2012aa}.
Its high gas mass would suggest that M31 has not undergone the interstellar medium ``blowout'' associated with quenching in recent cosmological hydrodynamic simulations.
Internal dynamics driven by structure in a galaxy can reduce the star formation efficiency of available gas (``bar quenching"; \citealt{Tubbs:1982aa,Khoperskov:2018vl,Newnham:2020tn}).
M31 is known to have a bar and so could be affected by this mechanism \citep{Athanassoula:2006vp,Dorman:2015vk,Feng:2022wr}.
The CGM would not be affected by the bar, allowing a star-forming galaxy level of \OVI\ with a low sSFR.

\subsection{Cool and hot gas in the CGM of quenched galaxies}
If \OVI\ around galaxies with $\Mstar=10^{10}\rng10^{11}$ \Msun\ is mostly collisionally ionized and found in $\sim 10^{5}$ K gas, then the \OVI\ dichotomy implies that quenching is associated with a drop in the amount of warm gas in a galaxy's CGM.
There is evidence that passive galaxies in the same mass range also have less $T\sim 10^{4}$ K gas than the corresponding star-forming galaxies.
Cool gas is traced by ions such as \specnotation{Mg}{II}.
Analyses of the covering fractions of strong \specnotation{Mg}{II} absorbers as a function of impact parameter, stellar mass, and star formation rate find that the covering fraction is several times lower near passive galaxies than near star-forming galaxies \citep{Bordoloi:2011vq,Lan:2020aa,Anand:2021aa}.

The situation is less clear for gas that is too hot to be traced by $\OVI$\ ($T\gtrsim 10^6$ K).
Emission from hot gas can be detected in X-rays.
\citet{Comparat:2022aa} and \citet{Chadayammuri:2022aa} use eROSITA data to measure X-ray emission around stellar-mass-controlled star-forming and passive galaxy samples and find conflicting results.
\citet{Comparat:2022aa} find that passive galaxies are associated with more X-ray emission while \citet{Chadayammuri:2022aa} find the opposite.
\citet{Chadayammuri:2022aa} argue that the discrepancy is driven by differences in how emission from groups and clusters is treated for galaxies other than the group and cluster centrals.
Determining which interpretation of the data is correct will tell us whether the CGM transformation associated with quenching mostly heats gas or also drives some of it out of the halo.
If warm gas is driven out as well as heated, the passive galaxy X-ray emission should be weaker.
If the gas is heated but remains in the CGM, then the passive galaxy X-ray emission should be stronger.

\section{Conclusion} \label{sec:conclusion}
We study the incidence rate of strong \OVI\ absorption in the circumgalactic medium of $z<0.6$ star-forming and passive central galaxies with stellar masses between $10^{10.1}$ and $10^{10.9}$ \Msun.
The galaxy-\OVI-absorber pair sample is drawn from \citet{Tchernyshyov:2022vw} and references therein and from the CUBS survey \citep{Chen:2020wv}.
The absorber impact parameters span $\approx 0$ to 400 physical kpc.
To separate differences due to galaxy stellar or halo mass from differences related to galaxy quenching, we compare the two galaxy classes in narrow stellar mass ranges, $\Mstar = 10^{10.1} \rng 10^{10.5}$ \Msun\ and $\Mstar = 10^{10.5} \rng 10^{10.9}$ \Msun, and, separately, in a relatively narrow estimated halo mass range, $\Mhalo \approx 10^{11.6}\rng 10^{12.2}$ \Msun.

For each combination of mass range and galaxy class, we further split the galaxies by impact parameter into an inner and outer region.
We measure covering fractions of sightlines with $\NOVI\geq10^{14.3}$ \cmmt\ for these two regions and calculate probabilities that the inner star-forming galaxy covering fractions are less than or equal to the inner passive galaxy covering fractions.
We also explore some possible origin scenarios for the small number of strong \OVI\ absorbers that are still found around passive galaxies.

Our observational results on the incidence rate of strong \OVI\ absorbers around star-forming and passive galaxies in the mass ranges $\Mstar = 10^{10.1} \rng 10^{10.5}$ \Msun, $\Mstar = 10^{10.5} \rng 10^{10.9}$ \Msun, and $\Mhalo \approx 10^{11.6}\rng 10^{12.2}$ \Msun\ are as follows:
\begin{itemize}
    \item Within 150 kpc, the covering fraction of strong \OVI\ absorbers is approximately $0.9\rng1$ for star-forming galaxies and $0\rng0.2$ for passive galaxies.
    \item In each mass range and within 150 kpc, the probability that star-forming galaxy covering fractions are less than or equal to passive galaxy covering fractions is less than 0.001. At greater than $3\sigma$-equivalent statistical significance, the incidence rate of strong \OVI\ absorption in the CGM is greater around star-forming galaxies than around passive galaxies with the same stellar or halo mass.
\end{itemize}

From these observational results, we reach the following conclusions:
\begin{itemize}
    \item There is a dichotomy in the incidence rate of strong \OVI\ absorption in the CGM of star-forming and passive $\Mstar \sim 10^{10}$ \Msun\ galaxies at fixed stellar mass and at fixed halo mass.
    \item The quenching of a $\Mstar \sim 10^{10}$ \Msun\ central galaxy at low redshift is, in most cases, accompanied by a transformation of the galaxy's CGM. This change is not driven by the mass of the galaxy's dark matter halo.
    \item There may be a delay between galaxy quenching and an observable change in the incidence rate of \OVI. Alternatively (or jointly), there may be a less common mode of quenching in which the CGM is not substantially changed.
\end{itemize}

\begin{acknowledgements}
KT thanks Alison Coil, Yakov Faerman, Chris McKee, Evan Schneider, Fakhri Zahedy, and Yong Zheng for useful discussions. KT, JKW, and MW acknowledge support for this work from NSF-AST 1812521 and NSF-CAREER 2044303. JKW acknowledges additional support as a Cottrell Scholar, from the Research Corporation for Science Advancement, grant ID number 26842. The authors gratefully acknowledge the UW Werk SQuAD (Student Quasar Absorption Diagnosticians), a team of more than 40 dedicated undergraduate researchers since 2017, who made significant contributions to the CGM$^2$ survey over the last five years and thus enabled some of the science presented in this work. This work benefited from lively Zoom discussions during KITP's ``Fundamentals of Gaseous Halos" program, and thus was supported in part by the National Science Foundation under Grant No. NSF PHY-1748958.
\end{acknowledgements}

\software{\texttt{astropy} \citep{Astropy-Collaboration:2013uv,Astropy-Collaboration:2018vm,Astropy-Collaboration:2022aa},
\\ \texttt{linetools} \citep{Prochaska:2017vh},
\\ \texttt{matplotlib}  \citep{Hunter:2007ux},
\\ \texttt{numpy} \citep{Harris:2020ti},
\\ \texttt{pandas} \citep{McKinney:2010vw}
}

\bibliography{main}{}
\bibliographystyle{aasjournal}

\end{document}